\documentclass[10pt,conference,letterpaper]{IEEEtran}
\pdfoutput=1

\usepackage[T1]{fontenc}
\usepackage[utf8]{inputenc}
\usepackage[english]{babel}

\usepackage{times}
\usepackage{balance}
\usepackage{csquotes}
\usepackage{microtype}

\usepackage{xcolor}
\usepackage[pdftex]{graphicx}

\usepackage{bm}
\usepackage{amsmath}
\usepackage{amsthm}
\usepackage{amsfonts}
\usepackage{amssymb}

\usepackage{siunitx}

\usepackage{url}
\usepackage{hyperref}

\usepackage[style=ieee]{biblatex} 
\usepackage[acronym,hyperfirst=false,nohypertypes={acronym}]{glossaries}
\usepackage{cleveref}

\usepackage[font=small,belowskip=-1.1em]{subcaption}

\usepackage{tabu}
\usepackage{booktabs}
\usepackage{xspace}

\usepackage[
        cachedir=.,
        frozencache=true,
    ]{minted}

\newcommand{\cpf}{\emph{DIgSILENT~PowerFactory}\xspace}
\newcommand{\call}{\emph{AIT Lablink}\xspace}

\newcommand{\clargo}{\emph{LarGo!}\xspace}
\newcommand{\cmosaik}{\emph{mosaik}\xspace}

\usepackage[disable]{todonotes}

\DeclareCaptionLabelSeparator{iaria}{.\quad}
\graphicspath{{./figures/}}
\DeclareGraphicsExtensions{.pdf,.jpeg,.png}

\newacronym[shortplural={CPS}]{CPS}{CPS}{cyber-physical system}
\newacronym{AI}{AI}{Artificial Intelligence}
\newacronym{AL}{AL}{Adversarial Learning}
\newacronym{ANN}{ANN}{Artificial Neural Network}
\newacronym{ARL}{ARL}{Adversarial Resilience Learning}
\newacronym{GAN}{GAN}{Generative Adversarial Network}
\newacronym{GPGPU}{GPGPU}{General-Purpose Graphics Processing Unit}
\newacronym{DL}{DL}{Deep Learning}
\newacronym{GRU}{GRU}{Gated Recurrent Unit}
\newacronym{ICT}{ICT}{Information and Communication Technology}
\newacronym{LSTM}{LSTM}{Long-Short Term Memory}
\newacronym{AEG}{AEG}{Attack Execution Graph}
\newacronym{ML}{ML}{Machine Learning}
\newacronym{PoC}{PoC}{Proof-of-Concept}
\newacronym{RL}{RL}{Reinforcement Learning}
\newacronym{DNN}{DNN}{Deep Neural Network}
\newacronym{RNN}{RNN}{Recurrent Neural Network}
\newacronym{CNN}{CNN}{Convolutional Neural Network}
\newacronym{SIMD}{SIMD}{Single Instruction, Multiple Data}
\newacronym{DoE}{DoE}{design of experiments}
\newacronym{DNC}{DNC}{Differentiable Neural Computer}
\newacronym{PV}{PV}{Photovoltaic}
\newacronym{STL}{STL}{Signal Temporal Logic}
\newacronym{MTL}{MTL}{Metric Temporal Logic}
\newacronym{MITL}{MITL}{Metric Interval Temporal Logic}
\newacronym{AV}{AV}{Autonomous Vehicle}
\newacronym[shortplural={MAS}]{MAS}{MAS}{Multi Agent System}
\newacronym{RTU}{RTU}{Remote Terminal Unit}
\newacronym{TCP}{TCP}{Transmission Control Protocol}
\newacronym{LPEP}{LPEP}{Lightweight Power Exchange Protocol}
\newacronym{FMI}{FMI}{Functional Mock-Up Interface}
\newacronym{XML}{XML}{eXtensible Markup Language}
\newacronym{AS}{AS}{Autonomous System}
\newacronym{QoS}{QoS}{Quality of Service}
\newacronym{VPN}{VPN}{Virtual Private Network}
\newacronym{IP}{IP}{Internet Protocol}
\newacronym{GIS}{GIS}{Geospatial Information System}
\newacronym{SIL}{SIL}{Software in the Loop}
\newacronym{HIL}{HIL}{Hardware in the Loop}
\newacronym{UDP}{UDP}{User Datagram Protocol}
\newacronym{JSON}{JSON}{Java-Script Object Notation}

\newglossaryentry{resilience}{
  name=resilience,
  description={The ability of a system to withstand unforseen, rare and
  potentially catastrophic events and to self-heal or improve in
  reaction to these events. Ideally resilience is increasing monotonously.}
}

\begin{document}
\title{\Large\bfseries%
    Large-Scale Co-Simulation of Power Grid and Communication Network Models\\
    with Software in the Loop}

\author{\centering\setlength{\tabcolsep}{0pt}%
  \begin{tabular}{cc}
  \multicolumn{2}{c}{Eric MSP Veith\textsuperscript{1}, Jawad Kazmi\textsuperscript{2}, and Stephan Balduin\textsuperscript{1}}\\[1ex]
  \parbox{0.5\linewidth}{\centering\textsuperscript{1}\,%
    OFFIS e.V.\\
    Oldenburg, Germany\\
    Email: \texttt{\{veith,balduin\}@offis.de}}
    & \parbox{0.5\linewidth}{\centering\textsuperscript{2}\,%
    Austrian Institute of Technology (AIT)\\
    Vienna, Austria\\
    Email: \texttt{jawad.kazmi@ait.ac.at}}
  \end{tabular}
}

\maketitle

\begin{abstract}

  Power grids are transitioning from an infrastructure model based on reactive
  electronics towards a smart grid that features complex software stacks with
  intelligent, pro-active and decentralized control. As the power grid
  infrastructure becomes a platform for software, so does the need for a
  reliable roll-out of software updates on a large scale. In order to validate
  resilient large-scale software roll-out protocols, corresponding test beds
  are needed, which mirror not only \gls{ICT} networks, but also include the
  actual software being deployed, and show the interaction between the power
  grid and the \gls{ICT} network during the roll-out, and especially during
  roll-out failures. In this paper, we describe the design implementation of a
  large-scale co-simulation test bed that combines \gls{ICT} and power grid
  simulators. We pay specific attention to the details of integrating
  containerized software in the simulation loop.

\end{abstract}


\section{Introduction}

The transition of the power grid to the smart grid is happening on a large
scale. From the first introduction of the term \emph{smart
grid}~\parencite{bush2014smartgrid}, assets in the power grid have evolved
into software platforms that feature a vast array of services. Transformers
have become tools in asset management~\parencite{7005528}, while \glspl{MAS}
represent nodes in the power grid~\parencite{veith2017universal}. 

The numerous use and business cases enabled by this kind of infrastructure
obviously require special attention to the software stack deployed on these
devices. The life and, hence, innovation cycle in the power grid of 30--60
years that was dominating in the traditional power grid does not hold anymore.
As the evolution of energy systems to \glspl{CPS} based on \gls{ICT}
technologies has happened, so has, with increased complexity, risen the
inherent risk of the overall system~\parencite{hanseth2007risk}. Since power
grids have become a target in terms of cyber security, as proven by the
attacks on the Ukrainian power grid between 2015 and
2017~\parencite{lee2016analysisukraine,prentice2017ukraine}. Specifically,
software solutions based on \gls{AI} technologies have been regarded as major
factors in technical debt, causing frequent updates to be
made~\parencite{Sculley2014,sculley2015hidden}.

In a recent literature survey, we noted that the emerging smart grid yields
numerous attack vectors, many stemming from the inclusion of \gls{ICT},
\gls{AI} technologies or tight market
integration~\parencite{veith2019cpsanalysis}. A major research gap exists in
\gls{AI}-based analysis of complex \glspl{CPS}, i.e., the combination of power
grid and \gls{ICT}. Specifically, the interaction of both components has
hitherto seldom been discussed. On this basis, \gls{ARL}---discussed
originally by \textcite{Fischer2019arl}---offers an approach \emph{based} on
\gls{AI} to explore any \gls{CPS} without domain-specific knowledge and find
weaknesses in its configuration. This can very well be applied to software
roll-out and update processes, too, provided a test bed for this exists.
Software update roll-outs are, for a simulation testbed, a special case, as
they require the actual software to be deployed within the simulation in order
to assess the impact of the roll out.

To this end, we present a co-simulation approach that features power grid,
\gls{ICT}, and software-in-the-loop simulators. We will detail the specific
development to facilitate a software-in-the-loop simulation on a large scale.
The rest of this paper is structured as follows: \Cref{sec:related-work}
provides context for this work. We will detail possible, generalized models
for our testbed in \Cref{sec:modelling}. We then offer insights into the
\gls{ICT} co-simulation in \Cref{sec:ict-sil-cosimulation}, which accounts for
a major portion of this paper. We discuss the overall development in
\Cref{sec:discussion}, and conclude with pointers to future work in
\Cref{sec:conclusion}.

\section{Related Work}
\label{sec:related-work}

Simulators for specific domains exist for many years now, drawing from the
standard rationale that, once the system and the interaction of its components
become too complex to describe them in terms of formulæ and automatons, a
simulation to assert assumptions is in order. For each individual domain, a
sound selection of simulators exist, such as \emph{pandapower} by
\textcite{pandapower.2018} and \emph{SIMONA} by \textcite{kittl2019large} for
power grids, or OMNeT++ by \textcite{10.5555/1416222.1416290} for \gls{ICT}
simulations.

However, to witness effects of the two domains interacting with each other,
none of the two is fully suited. Specifically when smart grid messaging is
considered---which is crucial to optimization protocols such as
COHDA~\parencite{hinrichs2014decentralized,niesse2017local} or
Winzent~\parencite{veith2013lightweight}---, this part of the simulation
becomes crucial. Previous simulation environments for testing smart grid
messaging have focused on other parts of the problem, such as using a
\gls{GIS} layer to model the feed-in of renewable energy
sources~\parencite{veith2014open}.

The combination of two or more simulators from different domains is
facilitated through \emph{co-simulation}. A co-simulator provides an
infrastructure to schedule, synchronize different simulators, and enable data
exchange between model instances run by the different simulators. One solution
is provided by the \emph{mosaik}
co-simulator~\parencite{steinbrink2019cpes}---the one, in fact, used to
developed the test bed presented in this paper---, other approaches to
co-simulation are employed, e.g., by
\emph{OpSim}~\parencite{drauz2018modular}, or
\emph{PTOLEMY~II}~\parencite{eker2003taming}. A co-simulation of power grid
and \gls{ICT} have been described by different authors using different pieces
of software~\parencite{hopkinson2006epochs,lin2012geco,kelley2015federated},
but without taking the question of software roll-out into account.

This paper introduces a smart grid software roll-out testbed, based on the
idea discussed by \textcite{kintzler2018large}. It details the reasoning
behind using a software-in-the-loop~(SIL)\glsadd{SIL} approach---namely, that
the software being rolled out itself is complex enough that an approximation
through models is not feasible. If the subject to the experiment, i.e., the
software, is abstracted away, the result of the roll-out protocol cannot be
validated.

\gls{SIL} co-simulation is not new. \textcite{8406023} use the \gls{SIL}
technique to validate railway controllers; real-time \gls{SIL} co-simulation
in the smart grid for performance measurements is done by, e.g.,
\textcite{bian2015real}. The OMNeT++ simulation environment offers facilities
for \gls{HIL} integration~\parencite{Wehner2017,kolsch2018hardware}. However,
when the software itself is subject to change in a co-simulation/\gls{SIL}
scenario, an extension needs to be developed to allow the integration of
changing, virtualized software containers. This research gap is addressed by
our solution.

\section{Modelling Power Grid and ICT}
\label{sec:modelling}

\begin{figure}
    \centering
    \includegraphics[width=1.0\linewidth]{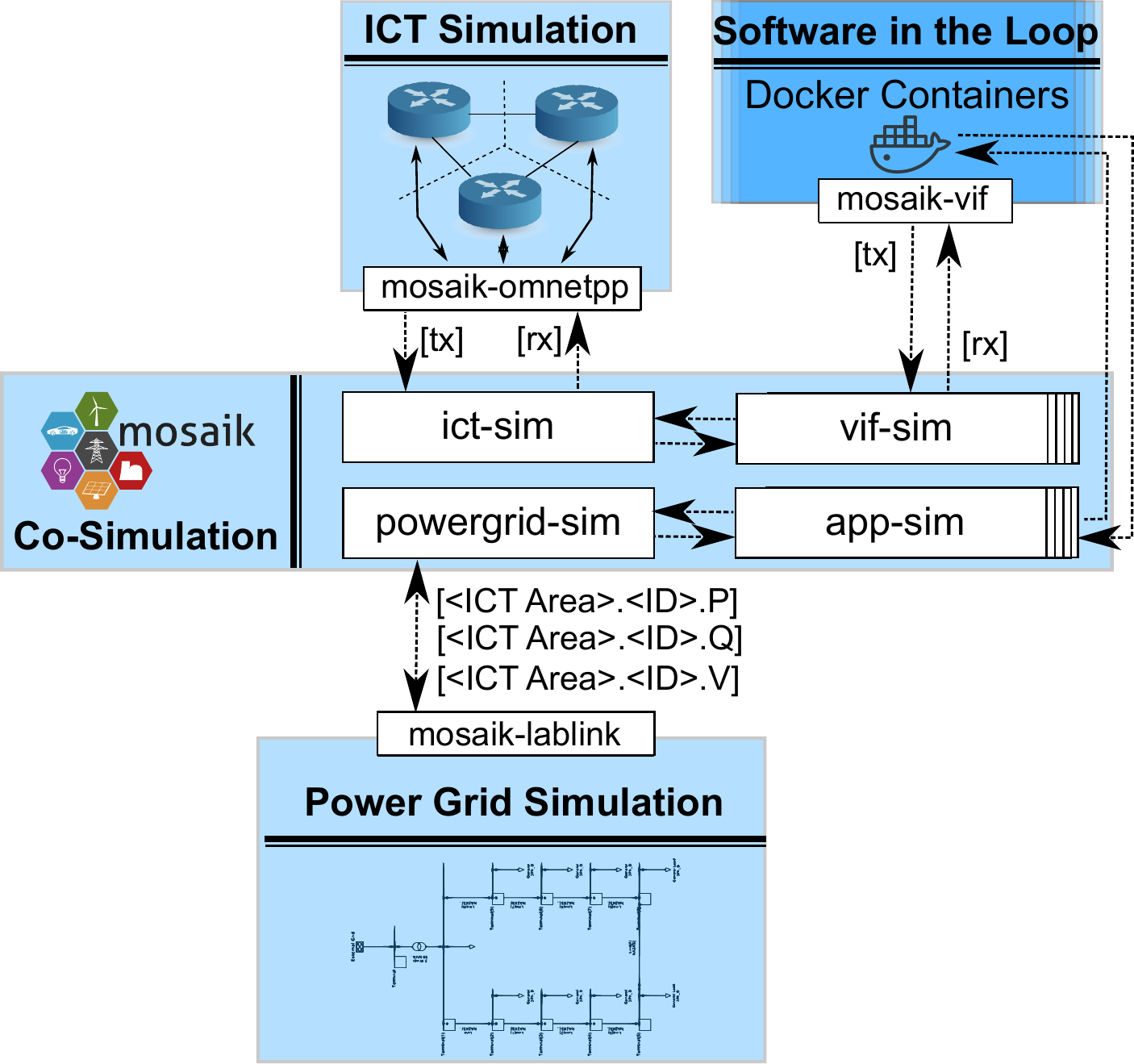}
    \caption{Data Exchange Schema of the Co-Simulation}
    \label{fig:cosimulation-data-exchange-schema}
\end{figure}

\Cref{fig:cosimulation-data-exchange-schema} shows the data exchange schema of
our co-simulation approach, including all software bridges that connect the
simulation with the \gls{SIL} part. The following sections will refer to the
schema when locating individual pieces of software.

\subsection{Power Grid Reference Model}

To capture the complex dynamics of and possible resulting effects caused by
the large-scale roll-out of smart power devices, a realistic and complete
model of the power system model is a necessity. The model needs to be detailed
that could later be simulated along with the other related components. It is
therefore important to choose a modeling and simulation tool that fulfills
these requirements. There are many good power system modeling and simulation
tools such as \emph{pandapower} by \textcite{pandapower.2018} and
\emph{SIMONA} by \textcite{kittl2019large} for power grids. After a survey and
discussion, \cpf was selected for the power system modeling as it meets the
selection criteria better than the other available tools. It provides a
detailed and fine-grained modeling and simulation of many aspects of the power
system are well known among the project partners and provides multiple
interface possibilities for doing a co-simulation.

\cpf is a sophisticated highly specialized, flexible and extendable platform
for power system modeling and simulation. It supports fine-grained power
system modeling and simulation through a combination of both the graphical and
scripting based methods for almost all the major areas of the power system
including generation, transmission, distribution etc. There is a large library
of models available that can be extended with writing custom components using
the DIgSILENT Simulation Language (DSL). For dynamic simulation of the power
system, the tool provides many functionalities including load and power flow
calculations, reliability and contingency analysis, RMS simulations and many
more. The tool also supports application programming interfaces (APIs) that
can be used to communicate with other simulators. It further supports the
automation using DIgSILENT Programming Language (DPL). 

\call is a multipurpose highly efficient, distributed, middleware for coupling
both hardware and software components in a co-simulation. It is used for
coupling the individual components and thus makes the power grid simulation
flexible and extendable.  \call provides interfaces, simulation control, and
data exchange capabilities. Using it, it is possible either doing simulations
or an emulation. It is developed and is extensively used in AIT for conducting
simulations for various research projects. A large set of hardware and
software components are already supported by bridges that make extending the
test-bed very easy.

Among the primary reasons for introducing \call into the co-simulation are
reuse and extendability. As mentioned previously, \call has been extensively
used in AIT and there are many software and hardware components that can be
readily included in the setup, if and when a need arises. This further enables
the reuse of this setup in other work packages of \clargo especially WP5 and
WP6 after adapting to the needs.

In the present setup depicted in figure here, \call provides
a message bus that the participating components (software/hardware) can
connect through a bridge. This bridge facilitates in the data exchange and
simulation control including the synchronization. The bridge and the
participating component have a one-to-one correspondence as indicated in the
figure. Two important such bridges are the \cpf and \cmosaik bridge. There are
some other \call system components like \textit{Synchronizer},
\textit{Simulation Manager} etc. that provide useful services but are excluded
here as they are part of every setup created with \call.

As the co-simulation is managed by \cmosaik, \call coordinates with \cmosaik
for data exchange and synchronization of the simulation. All the data exchange
request received from the coupled systems through \cmosaik are forwarded to
the respected component (\cpf in this case) while simulation synchronization
requests are forwarded to \textit{Synchronizer} that takes the appropriate
action.


\subsection{The Communications Infrastructure Model}

The \gls{ICT}~model serves to provide a number of realistic network areas to
dest the software roll-out scenarios. It is independent from the test bed
software, i.e., it was developed in parallel as part of the test bed, but can
be used on its own, e.g., without software in the loop. It features a number
of subnets, with each subnet area designating a certain characteristic network
environment, such as a well-built fibre channel network or a spotty wireless
area. To this end, it models an \gls{AS} with routers and intra-\gls{AS}
traffic/routing. These subnets have real IPv4 addresses assigned, as the
\gls{ICT}~model needs to process actual \gls{IP} traffic generated by the
existing software. The \gls{ICT}~infrastructure network is fully contained in
the class~C subnet

\begin{center}
  \texttt{10.64.0.0/10}~.
\end{center}

\Cref{tab:IctModel/Subnets} contains the relevant subnet specifications for
the areas that are described in the following paragraphs.

The reason for choosing this particular kind of subnet is its rather
remarkable subnet range and the fact that \texttt{10.64.0.0/10} is seldom used
as an IPv4 address space. This way, the \gls{ICT}~model does not collide with
existing private, class~C IPv4 addresses, such as those assigned by \gls{VPN}
software.

This leaves room for 8192~subnets with 254~hosts each in every defined
network. The \texttt{/24}-subnet should be the only network size, regardless
of how many hosts are contained in it. Routers always get the lowest IP
addresses assigned, i.e., \texttt{.1}, \texttt{.2}, \texttt{.3}, etc., before
the first hosts are added.

The test bed consists of 3~areas, which differ by their \gls{QoS} parameters.
We assume that most visible traffic we consider is either based on the
\gls{TCP} or employs similar mechanisms. This especially means that the
protocol features a retransmission algorithm. Since packet loss can be caused
either by a low-quality link or by network congestion, delay (denoted by
\(d\)) is the most describing parameter of a link (besides its data rate).

The first area is the \emph{Dedicated Network Area}. The underlying assumption
is that of the best possible infrastructure, where a grid operator has
deployed dedicated \gls{ICT}~cabling. Thus, the network is of high quality.
This does not only create a realistic scenario, but also serves as the test
case for the whole simulation infrastructure. The assumed nominal data rate is
\SI{1}{GBit \per s}; the delay is modeled stochastically per packet as:

\begin{equation}
  d \sim 10 + 50 \cdot f_{\lambda}(x, 1) ~ \text{[ms]}~.
\end{equation}

The function \(f_{\lambda}(x, 1)\) denotes the drawing of a random number from
an exponential distribution.

The second designated area is the \emph{Shared Links Area}. Here, we assume
that a grid operator uses the public infrastructure, such as internet-facing
connections. While we can assume that the necessary security precautions are
taken (e.g., by deploying a \gls{VPN} solution and generally encrypting
traffic), other traffic interferes with the \gls{QoS} of the update traffic we
examine. I.e., we can assume that there are occasional packet drops due to
congestion. As such, we model the delay as the drawing of a random number from
a normal distribution:

\begin{equation}
  d \sim \mathcal{N}(250; 20) ~ \text{[ms]}~.
\end{equation}

The area is well suited for variable-situation test cases. The link data rate
is still good, being at \SI{1}{GBit \per s} nominally.

The extreme end of the spectrum is modeled by the \emph{High-Impairment Area}.
It features low-datarate links (configurable from \SI{50}{kBit \per s} up to
\SI{100}{MBit \per s} with frequent congestion. This area also models the
deployment of wireless connections, such as 4G/CDMA~450 or similar
technologies. It is characteristic for an area where the development of the
infrastructure was hindered by, e.g., existing building situations, harsh
terrain, cost constraints, etc. As such, there are frequent packet drops and
even connection drops. The delay is modeled as:

\begin{equation}
  d \sim \mathcal{U}[100; \infty] ~ \text{[ms]}~,
\end{equation}

\noindent i.e., the drawing of a random number from a uniform distribution
with the interval \([100; \infty]\) (inclusive). A delay of infinity means the
link is broken.

\begin{table}
  \centering
  \caption{ICT model subnet specification}
  \label{tab:IctModel/Subnets}
  \begin{tabu}{XX}
    \toprule
    Network & 10.64.0.0/10 \\
    Network Range & 10.64.0.1 – 10.127.255.255 \\
    Dedicated Network & 10.64.0.0/12 \\
    Shared Links Network & 10.80.0.0/12\\
    High-Impairment Network & 10.96.0.0/12 \\
    Misc./Unallocated & 10.112.0.0/12 \\
    \bottomrule
  \end{tabu}
\end{table}

\section{ICT and Software-in-the-Loop Development}
\label{sec:ict-sil-cosimulation}

\subsection{Data Exchange Flows}

All simulators appear in the system twice, as
\cref{fig:cosimulation-data-exchange-schema} suggests. For each simulator
process---like the \gls{ICT} simulation, the power grid simulation, or each
containerized \gls{SIL}---also exists a representation as an entity object in
\emph{mosaik}. This object is responsible for connection data exchange
channels as well as communicating with the simulator processes. Overall, there
are at least four simulator processes with corresponding entity objects

The \textbf{\gls{ICT} Simulation} is responsible to run the communication
network simulation. Some nodes are also providing an interface to the
co-simulation as a bridge between the \gls{ICT} simulation and the \gls{SIL}
components. I.e., it also injects real \gls{IP} packets from the containerized
applications into the simulation environment and reads packets received from
other simulated nodes and transfers them back to the software containers. In
\cref{fig:cosimulation-data-exchange-schema}, it is represented as the
\texttt{ict-sim} object during a \emph{mosaik} run.
  
The \textbf{Power Grid Simulation} is responsible for calculating load flows
and line loads. It receives data through \emph{mosaik} from the actual
applications. E.g., an application representing an intelligent substation
would appear in the power grid simulation as substation; the substation
software would receive readings from the power grid simulation and issue
setpoints to it. This data flow is depicted in
\cref{fig:cosimulation-data-exchange-schema} as an exchange between the
\gls{SIL} container, the \texttt{app-sim} entity objects, and the
\texttt{powergrid-sim}.

The \textbf{application simulators} each represent one containerized piece of
software. They are not simulators in the strict sense, but the \gls{SIL}
component. The simulator is responsible for starting and stopping the
containers gracefully, and also for setting and collecting data coming from
the co-simulation or going to another simulator. Each application container
has its own application simulator and, hence, a corresponding \texttt{app-sim}
entity object.

Application logic will dictate communication with other containers. E.g., a
distributed real power schedule optimization heuristic like \emph{Winzent}
works on \gls{MAS} basis, and, therefore, requires communication with other
applications. I.e., the application containers are the logical connection
between \gls{ICT} and the power grid simulation. For the roll-out scenario, it
is not sensible to modify the application software to be part of the \gls{ICT}
simulation directly. Thus, we leave the applications in the container
undisturbed, and deploy a virtual network interface to connect the
applications to the \gls{ICT} simulation. This virtual network interface,
called \emph{vif} for short, also has a corresponding \texttt{mosaik-vif}
entity object in the \emph{mosaik} process. Thus, in our scenario, there
exist exactly as many \texttt{app-sim} object as there are \texttt{vif-sim}
entities.

The connection between the virtual interfaces and the corresponding nodes in
the is done in \emph{mosaik}. Any \gls{ICT}-related simulator offers at least
one model that represents the respective node. These models have exactly two
attributes, \texttt{rx} (``receive'') and \texttt{tx} (``transmit'').
\emph{Attribute} is a \emph{mosaik} term that designates a data exchange
interface for a simulator. Referring back to
\cref{fig:cosimulation-data-exchange-schema}, we see that each
\texttt{vif-sim} has these two attributes. The \texttt{rx} attributes always
receive data from \emph{mosaik}, whereas \texttt{tx} attributes transmit data
to \emph{mosaik}. The \gls{ICT} simulation has more than on
\texttt{tx}/\texttt{rx} pair: One for every node in the simulation for which a
corresponding application container exists. 

The connection in \emph{mosaik} is done in code like this:

\begin{minted}{python}
someapp_vif_entity = \
    someapp_vif_simulator.vif()
someapp_ict_entity = next(
    x for x in ict_model.children
    if x.eid == \
        'SimulatedNetwork/SomeApp/app-0')

world.connect(someapp_vif_entity,
              someapp_ict_entity,
              ('tx', 'rx'))
world.connect(someapp_ict_entity,
              someapp_vif_entity,
              ('tx', 'rx'),
              time_shifted=True,
              initial_data={'tx': None})
\end{minted}

\subsection{Virtual Network Adapter \& Packet Injection}

\begin{figure*}
  \centering
  \hspace{3ex}\begin{minipage}[t]{0.45\linewidth}
    \includegraphics[width=\textwidth]{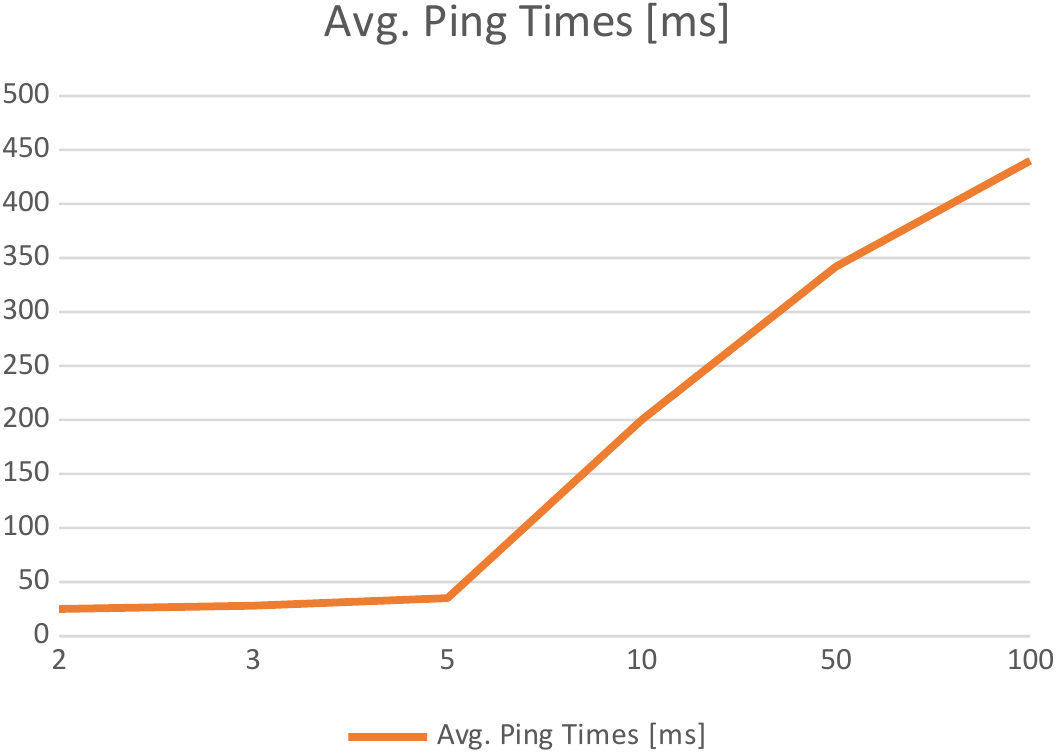}
  \end{minipage}\hfill\begin{minipage}[t]{0.45\linewidth}
    \includegraphics[width=\textwidth]{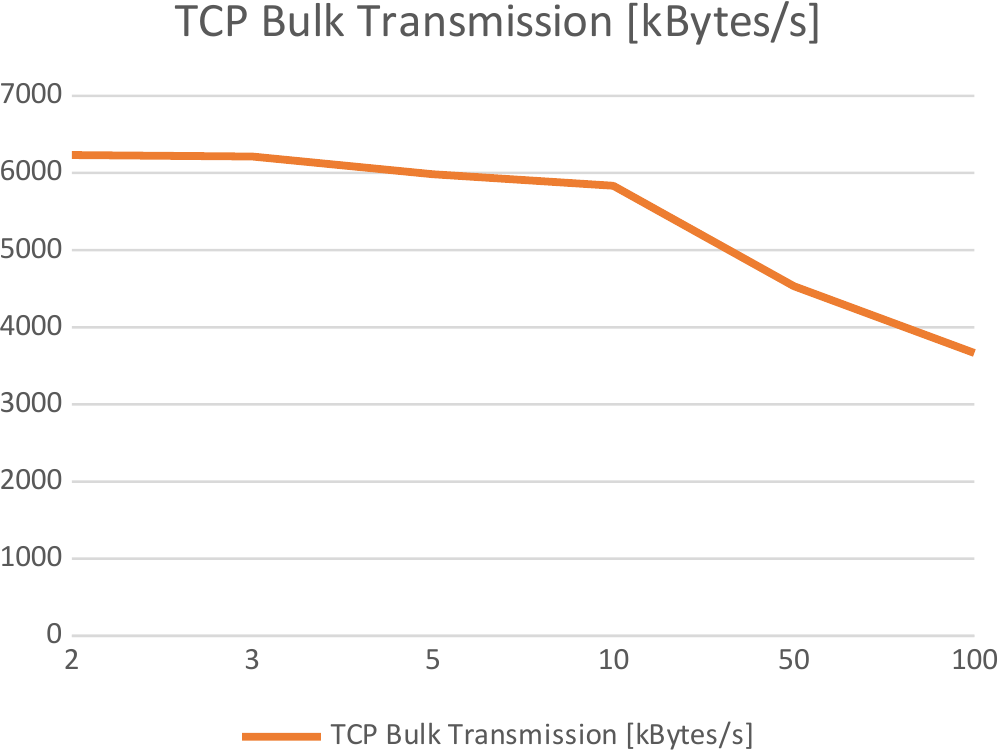}
  \end{minipage}\hspace{3ex}
  \caption{Experimental Measurements of Average Ping Times and Throughput}
  \label{fig:ping-throughput}
\end{figure*}

The code example in the previous section also shows how the hierarchical
addressing for entities in simulators in \emph{mosaik} is done. The
\texttt{vif} entities here denote a \gls{SIL} entity, i.e., a container with
a unmodified piece of software. Each entity denotes two software instances:
First, the virtual interface \emph{vif} that exists in a container, and the
\emph{vif-sim} that translates data between the container's networking stack
and the \emph{mosaik} co-simulation protocol.

These two pieces of software must exists separately as to avoid timing issues.
The startup behavior of container and its software cannot be observed by the
simulator; there exists a natural delay between launching the container and
being actually able to integrate it in the simulation run, i.e., the
containerized application sending and being able to receive packets. Since
multiple containers will normally be started, there is a time gap between the
first container's application being online and the last one being ready. As
\gls{SIL} implies no modification on the software, we cannot signal these
applications to hold until the simulation is ready to be started; hence, each
\emph{vif} must transparently buffer all data until the \emph{vif-sim} is
launched by the co-simulator.
In general, the \emph{vif} must act as if it was just a standard network
device. For this reason, the Linux kernel's tun/tap device driver was chosen.
It establishes a \emph{tun device} that appears as \texttt{tun0} (or any higher
index number) in the output of \texttt{ip address show}, can carry \gls{IP}v4
and \gls{IP}v6 addresses, and can be the subject of the default route.
Moreover, the tun device needs no gateway address, i.e., \texttt{ip route add
default dev tun0} without a \texttt{via} stanza is possible. This way, the tun
device transparently receives all traffic from the application, which does not
need to be changed; the kernel delivers all this traffic to a user space
application, i.e., the \emph{vif}. Injecting traffic is done the same way.
Since the userspace application needs to transmit this data to the
co-simulator, a second, specific rule for the \gls{IP} address of
the \emph{mosaik} instance is added, so that traffic between the simulator and
the \emph{vif} still flows via the standard \texttt{eth0} device.

As the tun device now tunnels all regular traffic, the communication protocol
between \emph{vif} and \emph{vif-sim} needs to be carefully chosen as to avoid
race conditions: Tunneling \gls{TCP} in \gls{TCP} is discouraged, as two
nested congestion control algorithms interfere with each other, creating
cascading time lags that can stall the application, known as \emph{\gls{TCP}
Meltdown}~\parencite{tcpmeltdown}. Since \gls{UDP} needs to be chosen, the
external address of the container is not known to the co-simulator, which
hinders the \gls{ICT} simulation from injecting data first before any data is
received from the container (and, thus, the container's address becomes
known). We solve this be simply sending a burst of zero \gls{UDP} `hello'
packets to announce the container.

Each byte of packet data received by the \emph{vif} is immediately transmitted
to the \emph{vif-sim}, which takes care of assembling the packets. Assuming
that the first transmission will contain the start of a \gls{IP} packet and no
intermediate packet data will be lost between container, \emph{vif} and
\emph{vif-sim}, the \emph{vif-sim} reads the packet length from the \gls{IP}
header field in order to assemble whole packets. These packets are then
encoded in Base64 format so that they can be transmitted to \emph{mosaik} via
\emph{mosaik's} \gls{JSON} communication protocol.

The \emph{vif-sim} as well as the \emph{mosaik-OMNeT++} adapter are
single-threaded, but use a cooperative, asynchronous I/O multitasking pattern
to handle the communication flow. Under the assumption that these applications
are I/O-heavy, but not computationally demanding, the single-threaded,
multi-process paradigm where much time is spent in the kernel's I/O space
suggests itself~\parencite{boostasio}.

\section{Discussion}
\label{sec:discussion}

As the general feasibility of co-simulation has already been established, we
focused prominently on the \gls{ICT} \gls{SIL} simulations. For this, we have
set up a co-simulation with a number of containers in which the
\emph{iPerf3}~\parencite{iperf3} application was running. We have deployed
pairs of clients and servers so that an iPerf client can send and receive data
from a dedicated iPerf server container. We used this set up to test both, the
average round-trip times (i.e., ICMP echo request/echo reply timings) and
\gls{TCP} bulk transfer speeds. All data was routed through the simulated
\gls{ICT} environment, so that the flow of data was as follows:
\emph{vif}---\emph{vif-sim}---\emph{mosaik}---\emph{OMNeT++}---\emph{mosaik}---\emph{vim-sim'}---\emph{vif'}.
The simulated \gls{ICT} environment does not impose additional artificial
delays in its network model.

\Cref{fig:ping-throughput} shows the behavior for both metrics given a rising
number of nods. Each data point represents a different number of nodes and the
average over 100~repeated simulation runs. Delays rise sharply as the number
of nodes rises, but not exponentially. With ping times in the area of
\SIrange{23}{447}{ms}, we assume that applications that do not realy on
real-time or, in general, low-latency communication can be accommodated by
this \gls{SIL} setup. However, the bulk throughput rate between
\SIrange{6102}{3654}{kB\per s} is far below a characteristic data rate
normally achieved by standard Ethernet connections.

We have investigated the reason for the low data rate and have identified
three major points. First, \emph{mosaik} currently uses non-compressed
\gls{JSON} messages in a request-reply pattern for data exchange with
out-of-process co-simulators. As both, the \emph{vif-sim}, and the
\emph{mosaik-OMNeT++} adapter, are written in C++, an additional network
round-trip is introduced, even if the simulation runs locally. In addition,
\emph{mosaik's} single-threaded request-response communication pattern with
its associated simulators means that dependent simulators expect an delay when
other simulators are being stepped or queried for data. 

Furthermore, \emph{mosaik} has currently no facilities to allow simulators to
signal the necessity to be stepped; simulator control is completely in the
hands of \emph{mosaik}. This means that \emph{mosaik} must poll all
\emph{vif-sims} as often as possible since the co-simulator has no other way
of knowing when data is available from a \gls{SIL} container. In contrast to
the \gls{ICT} simulation, data from applications arrives non-deterministic.
In general, we have observed delays in message processing stemming from the
context switches between kernel space and user space that frequently occur as
data from the containerized applications travel through several network
stacks.

Moreover, we currently launch one \emph{vim-sim} process per container, as
this is the easiest way from a software engineering organization perspective.
However, this means a separate \gls{TCP} connection per container, a new
process, and a new data stream. We therefore plan to implement a multiplexing
architecture in the \emph{vim-sim} part in order to reduce the number of
processes, and, hence reduce task and context switches.

We believe that this approach offers great flexibility and ease in modelling
\gls{ICT} networks with \gls{SIL}. As the development of \emph{mosaik} is
open source and already aimed at providing higher throughputs and lower delay
in the communication with external simulators---e.g., a ZeroMQ implementation
to replace the socket API already exists---, and the co-simulator is extended
to allow for event-discrete, non-deterministic simulators as they exist in
this scenario, we see an increase in the throughput in the near future.

\section{Conclusion \& Future Work}
\label{sec:conclusion}

In this paper, we have detailed requirements and issues encountered in a
\gls{SIL} co-simulation of software roll-outs in the power grid. We have shown
how an interaction of \gls{ICT} and the power grid can be simulated and how
complete containerized software stacks can be embedded into this
co-simulation.

In the future, we expect optimizations on implementation level, e.g., more
efficient transports and serialization techniques, as well as implementing
zero-copy primitives to reduce the number of copy operations and context
switches. On a broader research perspective, we expect that abstracting parts
of the system through surrogate models~\parencite{balduin2018surrogate} will
provide for a way to simulate large-scale roll-out procedures.

\section*{Acknowledgements}

The presented work is conducted in the framework of the joint programming
initiative ERA-Net Smart Grids Plus~\parencite{eranet-smart-grids-plus}, with
support from the European Union’s Horizon 2020 research and innovation
programme. On national level, the work was funded and supported by the
Austrian Climate and Energy Fund (KLIEN, ref.~857570), by German BMWi
(FKZ~0350012A), and by the Swedish Energy Agency (Project number 42794-1).

\printbibliography
\end{document}